\documentstyle[12pt]{article}

\def\ps@myheadings{\let\@mkboth\@gobbletwo
 \def\@oddhead{{\sl\rightmark}\hfil } %
 \def\@oddfoot{}\def\@evenhead{\rm \thepage\hfil\sl\leftmark} %
 \def\@evenfoot{}\def\sectionmark##1{}\def\subsectionmark##1 {}
 }

\newcommand{\sint}{\sin \theta}        % sin theta
\newcommand{\cost}{\cos \theta}        % cos theta
\newcommand{\nue}{\nu_{\rm e}}         % nu-e
\newcommand{\numyu}{\nu_{\mu}}         % nu-myu
\def\PTP{\em Prog.Theor.Phys.}
\def\PR{\em Phys.Rev.}
\def\PRL{\em Phys.Rev.Letters}
\def\SOKEN{\em Soryushiron Kenkyu} 
\def\PROC{\em Meeting on Beta Decays and Muon Captures in Complex Nuclei}   
\def\JETP{\em J.Exptl.Theoret.Phys. (U.S.S.R)}
\begin{document}
\title{BIRTH OF NEUTRINO OSCILLATION \thanks{Opening address at the Europhysics 
NEUTRINO OSCILLATION WORKSHOP (NOW' 98), 7-9 Sept. 1998, Amsterdam; MEIJO Preprint 24 Sept. 1998}}
\vskip 2em
\author{M. NAKAGAWA  \\
Department of Physics, Meijo University, Nagoya 468-8502, Japan \\
E-mail: mnkgw@meijo-u.ac.jp}

\date{}
\maketitle
\begin{abstract} 
 This is a brief presentation of historical introduction to the theoretical concept of 
neutrino oscillation during the early stage of the studies up to 60's.
\end{abstract}

\section{Prediction of $\mu$ and $\nu_{\mu}$ (1942) and Critiques by Sakata } \

  Let me begin with Sakata, that would concern with the theme of this talk. In 1942, Sakata 
and Inoue proposed the so-called "two-meson theory" ~\cite{S1}. Their theory was to claim 
the existence of another pair of leptons:
\begin{equation}
  {\rm m}^{-}(= \mu^{-}), \;\;\; {\rm n} (= \numyu), 
\end{equation}
in addition to ${\rm e}^{-}$ and $\nue$. Original meanig of the two-meson theory was the 
introduction of $\mu$ (called as  mu-meson) in addition to the Yukawa meson (pi-meson). It 
is to be noted that the 'neutrino' n was assumed as a particle different from $\nu (= \nue)$, 
and was not necessarily considered as a massless particle. However, the masses of the neutrinos 
were found to be very small, and afterwards the neutrinos were taken practically as massless 
particles and more-over as identical each other ($\nue = \numyu$) from the convenience and 
economy principles. 

  Sakata had been always sceptical to these conventional assumptions and repeatedly warned us 
that the principles of convenience and economy were dangerous and often missleaded physicists. 

  I would like to present another critique by Sakata. In 1955, Sakata wrote a paper in Japanese with 
a shocking title '{\it Superstition around Majorana Neutrino}'~\cite{S1}.  His claim is that it is not 
adequate to ask whether a neutrino be a Dirac or Majorana particle in alternatives, and it will be 
nothing but a superstition to believe that the answer be obtained by the experiment of the double beta 
decay and so on. His point is that such a question arises from choosing the interactions within 
unnecessarily small variety.  Two sorts of interactions for the beta decay are introduced with a 
Dirac neutrino $\nu$ as 
\begin{equation}
  H = \sum_{\rm i} G_{\rm i}(\overline{\Psi}_{\rm p}O_{\rm i}\Psi_{\rm n})
(\overline{\Psi}_{\rm e}O_{\rm i}\Psi_{\nu}) + {\rm h.c.},
\end{equation}
and
\begin{equation}
  H^{\rm c} = \sum_{\rm i} G_{\rm i}^{'}(\overline{\Psi}_{\rm p}O_{\rm i}\Psi_{\rm n})
(\overline{\Psi}_{\rm e}O_{\rm i}\Psi_{\nu}^{\rm c}) + {\rm h.c.},
\end{equation}
where $\Psi_{\nu}^{\rm c}$ means a charge conjugated field of $\Psi_{\nu}$.  As far as the neutrino is 
massless, these interactions are completely equivalent each other, and a Dirac neutrino can be defined 
from either of interactions just as a matter of naming. If, however, the beta interaction consists 
generally of both interactions as 
\begin{eqnarray}
 H_\beta &=& H + H^{\rm c}                        \nonumber  \\
 &=& \sum_{\rm i} (\overline{\Psi}_{\rm p}O_{\rm i}\Psi_{\rm n})
\{\overline{\Psi}_{\rm e}O_{\rm i}(G_{\rm i}\Psi_{\nu}+G_{\rm i}^{'}\Psi_{\nu}^{\rm c})\} + {\rm h.c.} \:,
\end{eqnarray}
the physical situation gets a drastic change.  When the coupling constants satisfy
\begin{equation}
 G_{\rm i} = G_{\rm i}^{'},        \label{Majo 1}
\end{equation}
the interaction can be transfered to the one defining a Majorana field.  However as a matter of experimental 
accuracy, a confirmation of neutrinoless double beta decay, if it is, would yield only a constraint on the 
coupling constants, say, to the extent as 
\begin{equation}
 G_{\rm i} \simeq G_{\rm i}^{'}.   \label{Majo 2}
\end{equation}
Conversely, eq.(\ref{Majo 2}) leads to the result almost equivalent to the one of the Majorana theory on 
the phenomenological level.  Furthermore, to confirm the Majorana theory literally, one must establish the 
equality eq.(\ref{Majo 1}) strictly for all the processes involving the neutrino.

  Sakata had on many occasions noted that people is used to believe in a simple idea on 'neutral' particles  
because of the difficulties of their detection, but exactly by this property the neutral particles would be 
gifted an unexpected nature.  His point is that the Majorana theory is equivalent to a specific choice of the 
neutrino interactions of the Dirac theory violating the lepton number conservation, and is to choose a specific 
(convenient and economical) type for the neutrino field beyond the accuracy of experimental confirmation. 
He, of cause, did not deny the theoretical significance of the Majorana particle.
We see such a case in an example of the two-component neutrino theory with only one chirality for the V-A weak 
interactions, which seems to have long been preventing people from considering the possibility of massive neutrinos. 

\section {Proposal of Neutrino Oscillation of a $\nu \leftrightarrow \overline{\nu}$ Transition (1957)}
\begin{flushright}
 B.Pontecorvo (1957)~\cite{P}
\end{flushright}

  We have another great physicist Pontecorvo, who had posed a strong question on the lepton number conservation. 
In 1957, Pontecorvo has indicated that if the two-component neutrino theory should turn out to be incorrect and 
if the conservation law of neutrino charge would not apply, then in principle neutrino $\to$ antineutrino 
transitions could take place in vacuo just on the analogy to the ${\rm K}^{0}$ to $\overline{\rm K}^{0}$ 
transition of Gell-Mann and Pais~\cite{GP}.

  In a subsequent paper(1958)~\cite{P}, Pontecorvo has put forth the idea to define the mixed particles as
\begin{eqnarray}
 \nu &=& \frac{1}{\sqrt{2}}(\nu_{1} + \nu_{2}),              \\
 \overline{\nu} &=& \frac{1}{\sqrt{2}}(\nu_{1} - \nu_{2}),   
\end{eqnarray}
where $\nu_{1}$ and $\nu_{2}$ are, he called, \it truely neutral Majorana particles \em which are mass 
eigenstates.  As important physical consequences, he pointed out that a stream of neutral leptons consisting 
mainly of antineutrinos when emitted from a nuclear reactor, will consists at some distance R from the reactor of 
\begin{equation}
 \overline{\nu} \to \overline{\nu}(50\%) + \nu (50\%),
\end{equation}
provided that either of $\nu_{1}$ or $\nu_{2}$ ceases out of the coherence leaving the other to survive.  So, 
this effect will cause a decrease of the capture cross-section of the antineutrinos to the half of the simple 
$\beta$ interaction, and the detection at different distances from the reactor will be needed.  And it was first 
pointed out the possibility of the observation of the oscillation effect on an astronomical scale, which has long 
been a key concept to solve the solar neutrino problem. 

\section{Proposal of Flavor Mixing and Flavor Oscillation of Neutrinos (1962)} 
\begin{flushright}
 Z.Maki, M.Nakagawa and S.Sakata (1962)~\cite{NAGOYA}    \\
  M.Nakagawa, H.Okonogi, S.Sakata and A.Toyoda (1963)~\cite{NAGOYA}
\end{flushright}

  The flavor mixing of neutrinos was proposed from a quite different line of thought than Pontecorvo's 
approach. It was based on an attempt at a unified understanding of leptons and hadrons.  After the proposal of 
the Sakata model of hadrons (1955)~\cite{S2}, it came to our attention that the weak interactions of the Sakata 
model (Okun'(1958)~\cite{OKN}) with a current, 
\begin{equation}
 J_\lambda = (\overline{e}\nu)_\lambda + (\overline{\mu}{\nu})_\lambda + (\overline{n}p)_\lambda + \epsilon(\overline{\Lambda}p)_\lambda,  \label{Okun}
\end{equation} 
have a lepton-baryon symmetry as follows;
\begin{equation}
 \nu,\;\; e, \;\; \mu \;\;\: \leftrightarrow \;\;\:  p,\;  n,\;  \Lambda \: (\:{\rm Sakata 
\;fundamental \;particles}\;),
\end{equation} 
provided the factor $\epsilon$ is assumed to be unit, which was pointed out by Gamba, Marshak and Okubo 
(1959)~\cite{GMO}. On the basis of this symmetry, the Sakata fundamental baryons were assumed as composite 
particles of leptons and a charged boson responsible for the strong interaction~\cite{MNOS}. 

   After the confirmation of the two-neutrino hypothesis~\cite{BRKHN}, we proposed a new model (Maki, Nakagawa 
and Sakata (1962))~\cite{NAGOYA} of the fundamental baryons modifying the above lepton-baryon symmetry to the 
correspondence of four leptons including two kinds of neutrinos.  In the model building we have assumed the 
following basic properties: \\
(1) Neutrinos should be of 4-component spinors in order to be seeds of the massive baryons.  Consequently, the 
neutrinos $\nu_{1}$ and $\nu_{2}$ to be bound in the baryons should have naturally their own masses.  We called 
these neutrinos as {\it true neutrinos}.   \\
(2) $\nue$ and $\numyu$ coupled to e and $\mu$ in the weak current should be mixing states of $\nu_{1}$ and 
$\nu_{2}$.  We called the neutrinos $\nue$ and $\numyu$ as {\it weak neutrinos}.  

  The mixing is expressed as  
\begin{eqnarray}
 \nue   &=& \cost \; \nu_{1} - \sint \; \nu_{2},  \nonumber  \\ 
 \numyu &=& \sint \; \nu_{1} + \cost \; \nu_{2},  \label{mixing}            
\end{eqnarray}
where we expressed the angle $\theta$ as $\delta$ in the paper, and the lepton-baryon correspondence
%%%%%%%%% Footnote {1} %%%%%%%%%%%%%%%
\footnote[1]{The correspondence was also proposed by Y.Katayama, K.Matumoto, S.Tanaka and E.Yamada, 
{\PTP} \underline{28}, 675 (1962) from a different point of view on neutrinos.}
%%%%%%%%% End of footnote [1} %%%%%%%%%%
are as follows: 
\begin{eqnarray}
 \nu_{1}   &\longleftrightarrow& \;\; p \nonumber           \\ 
 \nu_{2}   &\longleftrightarrow & \; X                      \\
 e^{-}     &\longleftrightarrow & \;\; n    \nonumber  \\                    
 \mu^{-}   &\longleftrightarrow & \;\; \Lambda .  \label{BLsymm}    \nonumber                  
\end{eqnarray}
In terms of the true neutrinos, the leptonic charged weak current is written as 
\begin{equation}
 j_\lambda = \cost(\overline{e}\nu_{1})_\lambda + \sint(\overline{\mu}{\nu}_{1})_\lambda - 
\sint(\overline{e}\nu_{2})_\lambda + \cost(\overline{\mu}\nu_{2})_\lambda,  \label{lepton}
\end{equation} 
and the baryonic charged weak current is obtained as 
\begin{equation}
 J_\lambda = \cost(\overline{n}p)_\lambda + \sint(\overline{\Lambda}p)_\lambda - 
\sint(\overline{n}X)_\lambda + \cost(\overline{\Lambda}X)_\lambda,    \label{baryon}
\end{equation}  
which reproduced the current suggested by Gell-Mann and L\'{e}vy~\cite{GL} modifying eq.(\ref{Okun})
as
\begin{equation}
  \frac{1}{\sqrt{1+\epsilon^{2}}}(\overline{n}p)_\lambda + 
    \frac{\epsilon}{\sqrt{1+\epsilon^{2}}}(\overline{\Lambda}p)_\lambda.
\end{equation}

  A few remarks should be added here: \\
(1) Sakata fundamental baryons are now taken to be quarks.   \\
(2) The structure of the baryonic weak charged current including mixing angle $\theta$ that we obtained is, 
when read in terms of the quarks, identical with the present quark current involving the Cabbibo angle 
that is transfered from the mixing angle of neutrinos. The proposal of the Cabibbo angle was made in 
1963~\cite{CAB}. \\
(3) As regards the origin of the mixing angle, we considered it as a realization of a mechanism making 
e and $\mu$ different, and attempted a simple model diagonalizing e and $\mu$ into the observed masses. 
However, the origin will be still one of the largest problems beyond the standard model.  \\
(4) The fourth baryon X came also naturally into the above correspondence.  But this particle was considered 
as having no seat in the weak current from unknown reason or as being a very large mass particle not yet 
discovered.   Later on, this particle became a candidate for the fourth quark, and was discovered as the 
charm~\cite{CHARM}. 

\subsection{Upper bound on the neutrino mass from the high energy neutrinos} 
\hspace*{15pt}Because of the particle mixing states of $\nu_{1}$ and $\nu_{2}$ with masses, the weak 
neutrinos $\nue$ and $\numyu$ are {\it not stable}\/ due to the the transmutation $\nue \leftrightarrow \numyu$. 
Therefore, we noted that a chain of reactions such as
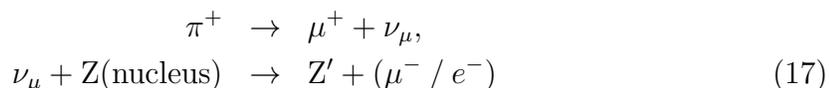
\begin{eqnarray}
  \pi^{+} &\to& \mu^{+} + \numyu ,     \nonumber \\
  \numyu + {\rm Z (nucleus)}\/ &\to& {\rm Z'}\/  + (\mu^{-} \: / \: e^{-})
\end{eqnarray}
will take place as a consequence of oscillation and will be only useful to check the two-neutrino hypothesis  
depending on the mass difference of $m_{1}$, $m_{2}$ which denote the masses of 
$\nu_{1}$ and $\nu_{2}$.  We defined the (half) oscillation time as 
\begin{eqnarray}
 T &=& \frac{\pi}{\vert E_{1} - E_{2}\vert}   \nonumber   \\
   &\simeq& 2\pi \frac{pc}{m_{2}c^2}\cdot \frac{M_p}{m_{2}}\cdot 0.7\times 10^{-24}{\rm sec} \:,  
\end{eqnarray}
where assumed as $m_{1} = 0$. 
%%%%%%%%%%%% Footnote[2] %%%%%%%%%%%%%%%%%
\footnote[2]{Here I present the formulas of the oscillation that we calculated at that time.  The calculation 
followed as a simple exercise to the ${\rm K}_{1}$ and ${\rm K}_{2}$ scheme of Gell-Mann and Pais(1955)~\cite{GP} 
but involving an arbitrary mixing angle as follows.

\noindent {\it Time development of ${\nu}_{\mu}$ from pion decay} \rm: 
\begin{eqnarray*}
 \vert {\nu}_{\mu},\:t \rangle = \Bigl\{e^{-iE_{1}t}\sin^{2}{\theta} + e^{-iE_{2}t}\cos^{2}{\theta}\Big\}\vert {\nu}_{\mu}\rangle + \frac{1}{2}\Bigl\{e^{-iE_{1}t} - e^{-iE_{2}t}\Big\}\sin2\theta \vert {\nu}_{e}\rangle.
\end{eqnarray*}

 Detection probability of ${\nu}_{e}$ at t : 
\begin{eqnarray*}
\vert \langle{\nu}_{e}\mid {\nu}_{\mu},\:t \rangle \vert^{2} = \frac{1}{2}\sin^{2}2\theta\{ 1 - \cos(E_{1} - E_{2})t\} .
\end{eqnarray*}

 Detection probability of ${\nu}_{\mu}$ at t : 
\begin{eqnarray*}
\vert \langle{\nu}_{\mu}\mid {\nu}_{\mu},\:t \rangle \vert^{2} = \sin^{4}\theta + \cos^{4}\theta + 
\frac{1}{2}\sin^{2}2\theta \cos(E_{1} - E_{2})t . 
\end{eqnarray*}   
\noindent {\it A half oscillation time of the detection probability of ${\nu}_{e}$ }\rm: 
\begin{eqnarray*}
 T = \frac{\pi}{\vert E_{1} - E_{2}\vert}.  
\end{eqnarray*}

 At relativistic limit, under an assumption $m_{2}\neq 0$, and $m_{1}\simeq 0$  \/,
\begin{eqnarray*}
 \vert E_{1} - E_{2}\vert &=& \vert \sqrt{{\bf p}^{2}+m_{1}^{2}} - \sqrt{{\bf p}^{2}+m_{2}^{2}}\; \vert \nonumber  \\
 &\simeq&  p( 1+ \frac{m_{2}^2}{2\, p^2} ) - p     \\
 &=&  \frac{m_{2}^2}{2\,p} \:.   
\end{eqnarray*}
\hspace*{10pt} Thus
\begin{eqnarray*}
 T &\simeq& \frac{2\pi p}{m_{2}^2}    \nonumber  \\ 
   & = & 2\pi \frac{pc}{m_{2}c^2}\cdot \frac{M_p}{m_{2}}\cdot 0.7\times 10^{-24}{\rm sec} \:, 
\end{eqnarray*}
\hspace*{12pt}where $M_p$ means the proton mass.   \\
\hspace*{12pt}See also, for example, A.K.Mann and H.Primakoff, \PR \em D\underline{15}, 655 (1977); S.M.Bilenky \\
\hspace*{12pt}and B.Pontecorvo, {\it Physics Report}, \underline{41}, 225 (1978).  }
%%%%%%%% End of footnote [2] %%%%%%%%%%%%%%%

    We have analyzed neutrinos of the famous experiment by Danby et al.(1962)~\cite{BRKHN} 
which confirmed the two-neutrino hypothesis.  Geometry of the neutrino path was taken as 100m, the flight time is 
\begin{equation}
 t_G = \frac{1}{3}\times 10^{-6}{\rm sec}. 
\end{equation}
 Assume for the neutrino beam as
\begin{eqnarray}
 pc &=& 1\:{\rm BeV},      \nonumber  \\
 m_{1}c^{2} &=& 0,                \\
 m_{2}c^{2} &=& x\:{\rm MeV}.  \nonumber
\end{eqnarray}
 Then no observation of $\nu_e$ would mean $ T \ge t_G $, which gives an upper bound 
\begin{equation}
m_{2}c^{2} \leq 3 \cdot 10^{-6}\: {\rm MeV}.
\end{equation} 

\subsection{Observation of electrons does not disprove the two-neutrino hypothesis in $\pi \to \mu +\nu_{\mu}$ chain}
\hspace*{15pt}We have also remarked an emission of a massive neutrino together with a massless neutrino would cause an 
apparent change in the magnitudes of the effective $\beta$ coupling constants depending on the Q-values 
and also would show an anomalous kink in the Kurie-plots as the threshold effect of massive neutrino. 
The $\beta$ interaction is now given from eqs.(\ref{lepton}), (\ref{baryon}) as 
\begin{equation}
  -L_{\beta} = \frac{G_{\rm F}}{\sqrt{2}}(\overline{p}n)_{\lambda}\{\cos^{2}\theta (\overline{e}\nu_{1})_\lambda 
 - \cos\theta\sin\theta(\overline{e}\nu_{2})_\lambda \} + {\rm h.c}.
\end{equation}
This means that the $\beta$-transition emitting $\nu_{1}$ is determined by an effective coupling constant  
$G_{\rm F}\cos^{2}\theta$ , but the $\nu_{2}$-emitting trnasition by that of $-G_{\rm F}\cos\theta\sin\theta$.  
Then when the Q-value is so small, the $\beta$ decay emits only the $\nu_{1}$ (in this analysis, we assumed 
$m_{1} \simeq 0 $), whereas the Q-value is large, the decay emits both of neutrinos 
%%%%%%%%%%%% Footnote{3} %%%%%%%%%%%%%%
\footnote[3] {We have used the following formula for the Kurie-plot analysis as
\begin{eqnarray*}
 \sqrt{\frac{N(E)}{G_{F}\cdot(G-T)pE}} = \left\{(E_{0}-E)^{2} + \epsilon\lbrack (E_{0}-E)^{2} - 
m_{2}^{2}\rbrack^{\frac{1}{2}}(E_{0}-E)\right\}^{\frac{1}{2}},  
\end{eqnarray*}
where
\begin{eqnarray*}
 \epsilon &=& 0 \quad \quad \quad {\rm for\; the\; emisson\; of\; only\; \nu_{1}}\/,     \\
          &=& \frac{\sin^{2}\theta}{\cos^{2}\theta} \quad {\rm for\; the\; emission\; of\; \nu_{1}\; and \; 
              \nu_{2}}\/. 
\end{eqnarray*}  }.
%%%%%%%%%% End of footnote {3} %%%%%%%%%%%

  To get the real information of the masses and mixing angle of neutrinos, we studied the data 
of nuclear $\beta$ decays.  Just in those times, there were reported an anomalous kink in the 
Kurie-plots by Langer group (we called this effect as "Langer effect") and also an increase of 
the magnitudes of the coupling constants with increase of the Q-values after subtraction of the 
radiative corrections~\cite{BETA}. These suggested \\
\begin{eqnarray}
 m_{2}c^{2} &\simeq& 1\:{\rm MeV},    \nonumber  \\
 \sin {\theta} &\simeq& 0.16 \sim 0.25.
\end{eqnarray}

   Under this mass condition, the $\cos(E_{1}-E_{2})t$ term vanishes in the probability oscillations of $\nue$ 
and $\numyu$, thus the ratio of $N_e$ to $N_{\mu}$ to be observed in the two-neutrino experiment initiating from 
$\pi \to \mu +\nu_{\mu}$ is given in terms of only the mixing angle $\theta$ as
\begin{eqnarray}
 \frac{N_e}{N_{\mu}} &=& \frac{2\sin^{2}\theta \cos^{2}\theta}{\cos^{4}\theta + \sin^{4}\theta}  \nonumber  \\
 &\simeq& \frac{1}{20} \sim \frac{1}{8} \: .
\end{eqnarray}
 We heard  Brookhaven had 29 $\mu^{-}$ with 8 showers found at that time. 

\subsection{Other processes} 
\hspace*{15pt} We also  computed the decay processes $\mu \to e + \gamma$ and $\nu_2 \to \nu_{1}+ \gamma$ 
with diagrams involving the weak boson; these can take place only through the muon-number non-conservation.  And 
we realized these diagrams can be given in terms of mass (squared) differences of virtual leptons on account of 
cancellations of divergent terms due to the rotation caused by the mixing angle (later on, this was called as 
G.I.M. mechanism~\cite{GIM}).  The decay amplitudes are controlled by factors $(m_{1}^{2} - 
m_{2}^{2})/M_{W}^{2}$ for $\mu \to e + \gamma$, and $(m_{\mu}^{2} - m_{e}^{2})/M_{W}^{2}$ for $\nu_2 \to 
\nu_{1}+ \gamma$ up to the Feynman integral factors
%%%%%%%%%%%% Footnote{4} %%%%%%%%%%%%%%
\footnote[4] {For the Feynman integral factor of the $\mu \to e + \gamma$ process, we followed the work of M.E.Ebel 
and F.J.Ernst, {\em Nuovo Cimento} \underline {15}, 173 (1960) by noting that $m_{2}$ plays their cutoff under 
$m_{1}=0$. And for the $\nu_2 \to \nu_{1}+ \gamma$, we calculated it keeping only the leading term.
}.
%%%%%%%%%% End of footnote {4} %%%%%%%%%%%
 The numerical results were
\begin{eqnarray}
 Br(\mu \to e + \gamma ) &\simeq&  10^{-17},      \nonumber   \\
 \tau(\nu_2 \to \nu_{1}+ \gamma) &\simeq&  10^{10}\:{\rm sec},
\end{eqnarray}
under the same parameters $m_{1} = 0, m_{2} = 1 {\rm MeV},  \sin {\theta} \simeq 0.16 \sim 0.25, 
M_{\rm W} = 1{\rm BeV}$ . 

\section {Flavor Oscillation of Majorana Neutrino (1967)} \
\begin{flushright}
 B.Pontecorvo (1967)~\cite{PG}   \\
 V.Gribov and B.Pontecorvo (1969)~\cite{PG}
\end{flushright}

   In 1967~\cite{PG}, Pontecorvo proposed the violation of the muon charge together with the violation of 
the leptonic charge conservation of the following type as 
\begin{equation}
  \nu \leftrightarrow \overline \nu \quad {\rm and} \quad  \nue \leftrightarrow \numyu.
\end{equation}
Again the transition of an active particle to a sterile particle takes place here as a transition el-neutrino 
$\leftrightarrow$ mu-neutrino, that is, physically as the flavor oscillation. \

   The above concept has been given a beautifull formulation by Gribov and Pontecorvo in 1969~\cite{PG}, 
which may be a first formulation, to my knowledge, of the Majorana mass terms of the Dirac neutrinos.  
Assumed neutrinos are the massless two-component Dirac neutrinos $\nu_{\rm eL}$ and $\nu_{\mu\rm L}$, which 
construct the weak interactions. The mass term of the Lagrangean is assumed as
\begin{equation}
 L_{\rm int} = m_{\rm e\overline{e}}\overline{(\nu_{\rm eL})^{c}}\nu_{\rm eL}
             + m_{\mu \overline{\mu}}\overline{(\nu_{\mu\rm L})^{c}}\nu_{\mu\rm L}
             + m_{\rm e\overline{\mu}}\overline{(\nu_{\rm eL})^{c}}\nu_{\mu\rm L} + {\rm h.c.},
\end{equation}
where e.g. $(\nu_{\rm eL})^{c}$ means a charge conjugated spinor.  Diagonalization of this Lagrangean 
leads to the Majorana particles $\phi_{1}$ and $\phi_{2}$ which have each eigenmass, and the original 
weak left-handed neutrinos are expressed as
\begin{eqnarray}
 \nu_{\rm eL} = \frac{1}{2}(1+\gamma_{5})(\phi_{1}\cos\xi + \phi_{2}\sin\xi),   \\
 \nu_{\mu\rm L} = \frac{1}{2}(1+\gamma_{5})(\phi_{1}\sin\xi - \phi_{2}\cos\xi),
\end{eqnarray}
where the mixing angle $\xi$ is determined in terms of $m_{\rm e\overline{\mu}}$, $m_{\rm e\overline{e}}$ 
and $m_{\mu\overline{\mu}}$. \\

\section {Summary}   \

   I have presented a brief historical introduction on the neutrino theory carried out in early stage up to 60's 
where the concept of the neutrino oscillation has born.  The developements of the theory after this stage are 
well known so that I would apologize for skipping the other many important contributions. 

   The motivation to the concept of the neutrino oscillation seems, to me, to consist in two main streams. 
One of them may be of a strong question on the conservation laws concerning leptonic charges either or both of 
the lepton number and the muon charge that would be violated just in analogy with the established evidence of 
${\rm K}^{0}$ to $\overline{\rm K}^{0}$ transition.   The other is in the attempt at model building for a 
unified understanding of the leptons and the fundamental entities of hadrons.  A unification of four leptons 
and the fundamental baryons at that time suggested the mixing scheme for neutrinos that explained
successfully the structure of the baryonic weak current; the universality of the weak interaction and the 
smallness of strangeness changing interaction.  

   The above features of intentions are, I would like to say, in principle still alive in the present stage of 
neutrino study as the quests for the origin of the mixing and for the true nature of the neutrinos.  I would hope 
the physics of leptons will be much deepened over every generation of flavors to open the new realm beyond the 
standard model, and indeed this workshop will remain as a great milestone for the future progress. @\\

\end{document}